\documentclass[twocolumn]{autart}

\usepackage{tikz}

\usepackage{amsmath}
\usepackage{latexsym, amssymb}
\usepackage{graphicx}
\usepackage{amsmath, amsbsy}
\usepackage{amsopn, amstext}
\usepackage{hyperref}
\usepackage{cancel, color}
\usepackage{epstopdf}

\def\J{{\bf 1}}

\DeclareMathOperator{\Span}{Span}
\DeclareMathOperator{\Col}{Col}
\DeclareMathOperator{\Row}{Row}

\DeclareMathOperator{\lcm}{lcm}
\DeclareMathOperator{\argmax}{argmax}
\def\cal{\mathcal}

\def\ra{\rightarrow}

\def\w{\wedge}

\def\d{\delta}

\def\D{\Delta}

\def\0{{\bf 0}}
\def\J{{\bf 1}}

\def\leqs{\leqslant}
\newcommand{\R}{{\mathbb R}}

\def\dsum{\mathop{\sum}\limits}
\def\thefootnote{*}

\newtheorem{dfn}[thm]{Definition}
\newtheorem{prp}[thm]{Proposition}
\newtheorem{exa}[thm]{Example}

\begin{document}

\begin{frontmatter}

\title{Weighted and Near Weighted Potential Games with Application to Game Theoretic Control\thanksref{footnoteinfo}}

\thanks[footnoteinfo]{This paper was not presented at any IFAC
meeting. This work is supported partly by NNSF
  62073315, 61074114, and 61273013 of China. Corresponding author: Daizhan Cheng. Tel.: +86 10 6265 1445; fax.: +86 10 6258 7343.}

\author[AMSS]{Daizhan Cheng}\ead{dcheng@iss.ac.cn},
\author[AMSS,CAS]{Zhengping Ji}\ead{jizhengping@amss.ac.cn}

\address[AMSS]{Key Laboratory of Systems and Control, Academy of Mathematics and Systems Science, Chinese Academy of Sciences, Beijing 100190, P.R.China}
\address[CAS]{School of Mathematical Sciences, University of Chinese Academy of Sciences, Beijing 100049, P.R.China}


\begin{abstract}
An algorithm is proposed to verify whether a finite game is a weighted potential game (WPG) without pre-knowledge on its weights. Then the algorithm is also applied to find the closest WPG for a given finite game. The concept and criterion of near weighted potential games (NWPGs) are given, indicating the evolutionary equivalence between a finite game and its closest WPG. Based on this, a method is proposed for replacing PGs by NWPGs in game theoretic control, which improved the applicability of PG-based optimization.
\end{abstract}

\end{frontmatter}

\section{Introduction}

In 1973, Rosenthal proposed a new type of games called potential games (PGs) in \cite{ros73}. The paper also revealed that some practically important games such as congestion games are potential. Since then many follow up investigations have been done \cite{har89,blu93,mil96,mon96a}. A systematic research on fundamental properties of PGs was presented in \cite{mon96b}. The theory of PGs has also been used to many engineering problems, including distributed power control and scheduling \cite{hei06}, road pricing \cite{wan13}, consensus of multi-agent systems \cite{mar09}, etc. Particularly, the fact that PGs have pure Nash equilibria makes the optimization via game-based approach technically possible, therefore it has become the kernel of game theoretic control \cite{gop11,gop14}.

Unfortunately, the set of finite PGs forms a lower dimensional subspace in the whole space of finite games \cite{can11,che16}. This essential weakness impedes its applications to many game-based systems. There are two ways to improve the applicability of PGs: (i) weighted potential game (WPG) approach \cite{mon96b}; (ii) near potential game (NPG) approach \cite{can13}.

Consider the WPG approach. Since a WPG has exactly the same advantages as a PG, and the set of WPGs is much larger than that of PGs, it is a reasonable idea to replace the former by the weighed ones to improve its applicability. However, there have been almost no applications of WPGs reported in literature, the obstacle for which lies in that there is no efficient way to determine a set of proper weights. As for a pre-assigned set of weights, the game is obviously equivalent to a PG; to the authors' best knowledge, there is no efficient way to verify if a finite game is a WPG, unless the weights are pre-assigned.

Next, consider the NPG approach. As is described in \cite{can13}, an NPG comes from a PG with certain small perturbations on its payoff functions. This kind of NPGs are of limited usages, because for a given game it is difficult to see whether there is a PG ``nearby". A more challenging problem is, it is very likely that there exists a WPG, which is closer than any other PG to the given game, but we were not able to find it up to now.

The first aim of this paper is to provide an algorithm to verify whether a finite game is a WPG, without assigning its weights in advance. Secondly, for a given finite game, the algorithm can provide the closest WPG to approximate it. Meanwhile, if the weights of a WPG are unknown, the algorithm can be used to reveal them. 
Hence, this method makes it possible to take WPG as a substitute for PG, paving a way for applying them to practical problems. For instance, the game theoretic control problem \cite{gop11} is considered, where a PG is the kernel for the control design. With the aid of evolutionary equivalence, a technique is proposed to replace PGs by near weighted potential games (NWPGs), which provides much more freedom for designing game-based optimization.

Our main tool is the semi-tensor product (STP) of matrices, which is a generalization of conventional matrix product \cite{che12a}.
It has been successfully applied to the analysis and control of Boolean networks \cite{che11,che12b,for13,las13}, graph theory \cite{wan12}, etc. Recently, it has also been used to investigate the problems of (networked) evolutionary games \cite{che15,guo13}.

The rest of this paper is organized as follows. In Section 2 we introduce some preliminaries, including (i) a brief review of STP; (ii) the definition of PGs;  (iii) the algebraic expression of finite games. Section 3 considers the weighted potential equation (WPE). Moreover, the formula for calculating potential functions by solving WPEs is presented. Some algebraic properties of WPEs, including its subspaces, dimension, and basis are revealed. Finally, an algorithm is obtained to calculate the weights of a WPG. Section 4 introduces the concept of NWPGs and an algorithm to check the evolutionary equivalence of a game with its closest WPG. In Section 5, a technique is provided to find the closest WPG for a given finite game while its potential function is pre-assigned. This technique makes it possible to substitute PGs with NWPGs in game theoretic control. Section 6 is a conclusion.

Before ending this section some notations are listed as follows.

\begin{enumerate}

\item  ${\cal M}_{m\times n}$: the set of $m\times n$ real matrices.

\item $\Col(M)$ ($\Row(M)$) is the set of columns (rows) of a matrix $M$. $\Col_i(M)$ ($\Row_i(M)$) is the $i$-th column (row) of $M$.

\item ${\cal D}_k:=\left\{1,2,\cdots,k\right\},\quad k\geqslant 2$.

\item $\d_n^i$: the $i$-th column of the identity matrix $I_n$.

\item $\D_n:=\left\{\d_n^i\vert i=1,\cdots,n\right\}$.

\item $\J_k={\underbrace{(1,1,\cdots,1)}_k}^T$.


\item A matrix $L\in {\cal M}_{m\times n}$ is called a logical matrix
if the columns of $L$ are of the form of $\d_m^k$. That is, $\Col(L)\subset \D_m$. Denote by ${\cal L}_{m\times n}$ the set of $m\times n$ logical
matrices.

\item If $L\in {\cal L}_{n\times r}$, by definition it can be expressed as
$L=[\d_n^{i_1},\d_n^{i_2},\cdots,\d_n^{i_r}]$. For brevity, it is denoted by $L=\d_n[i_1,i_2,\cdots,i_r]$.


\end{enumerate}

\section{Preliminaries}

\subsection{Semi-tensor Product of Matrices}

This subsection provides a brief survey on semi-tensor product (STP) of matrices. We refer to \cite{che12a,che12b} for more details.
\begin{dfn} \label{d2.1.1} \cite{che11,che12a}:
Let $M\in {\cal M}_{m\times n}$, $N\in {\cal M}_{p\times q}$, and $t=\lcm\{n,p\}$ be the least common multiple of $n$ and $p$.
The STP of $M$ and $N$, denoted by $M\ltimes N$, is defined as
\begin{align}\label{2.1.1}
M\ltimes N:=\left(M\otimes I_{t/n}\right)\left(N\otimes I_{t/p}\right)\in {\cal M}_{mt/n\times qt/p},
\end{align}
where $\otimes$ is the Kronecker product.
\end{dfn}

We briefly review some basic properties:

\begin{prp}\label{p2.1.3}
\begin{enumerate}
\item (Associative Law) $\forall F\in {\cal M}_{m\times n}$, $G\in {\cal M}_{p\times q}$, $H\in {\cal M}_{r\times s}$,
\begin{align}\label{2.1.2}
(F\ltimes G)\ltimes H = F\ltimes (G\ltimes H).
\end{align}
\item (Distributive Law) $\forall a,b\in \R$,
\begin{align}\label{2.1.3}
\begin{cases}
F\ltimes (aG\pm bH)=aF\ltimes G\pm bF\ltimes H,\\
(aF\pm bG)\ltimes H=a F\ltimes H \pm bG\ltimes H.
\end{cases}
\end{align}
\end{enumerate}
\end{prp}
As for the transpose and inverse, we have
\begin{prp}\label{p2.1.7}
\begin{align}
\label{2.1.8} (A\ltimes B)^\mathrm{T}=B^\mathrm{T}\ltimes A^\mathrm{T}.
\end{align}
\end{prp}
\begin{prp}\label{p2.1.8}
Assume $A$ and $B$ are invertible, then
\begin{align} \label{2.1.9}
(A\ltimes B)^{-1}=B^{-1}\ltimes A^{-1}.
\end{align}
\end{prp}

\begin{rem}\label{r2.1.2}
	Note that when $n=p$, $M\ltimes N=MN$. That is to say, the STP is a generalization of conventional matrix product. Moreover, it keeps almost all the properties of conventional matrix product \cite{che12a}. Hence throughout this paper the matrix product is assumed to be the STP and the symbol $\ltimes$ is mostly omitted.
\end{rem}

The following property is for STP only.

\begin{prp}\label{p2.1.10} Let $X\in \R^m$ be a column vector and $M$ a matrix. Then
\begin{align}\label{2.1.11}
X\ltimes M=\left(I_m\otimes M\right)X.
\end{align}
\end{prp}


\begin{dfn}\label{d2.1.12} \cite{che11,che12a} A matrix $W_{[m,n]}\in {\cal M}_{mn\times mn}$, defined by
\begin{align}\label{2.1.13}
W_{[m,n]}=\left[I_n\otimes \d_m^1, I_n\otimes \d_m^2,\cdots,I_n\otimes \d_m^m \right].
\end{align}
is called the $(m,n)$-th swap matrix.
\end{dfn}


The basic function of the swap matrix is to ``swap" two vectors. That is,

\begin{prp}\label{p2.1.14} Let $X\in \R^m$ and $Y\in \R^n$ be two columns. Then
\begin{align}\label{2.1.15}
W_{[m,n]}\ltimes X\ltimes Y=Y\ltimes X.
\end{align}
\end{prp}

%
%
%
%

\subsection{From Finite Game to Finite Potential Game}

\begin{dfn}\label{d2.2.1} \cite{gib92} A finite (non-cooperative normal)  game is a triple $G=(N,S,C)$, where
\begin{enumerate}
\item[(i)] $N=\{1,2,\cdots,n\}$ is the set of players;
\item[(ii)] $S=\prod_{i=1}^nS_i$ is called the profile, where $S_i=\{1,2,\cdots,k_i\}$, $i=1,\cdots,n$ is called the set of strategies of player $i$.
\item[(iii)] $C=(c_1,\cdots,c_n):S\ra\R^n$, where each component $c_i:S\ra \R$ is called the payoff (or utility) function of player $i$.
\end{enumerate}

Denote by ${\cal G}_{[n;k_1,\cdots,k_n]}$ the set of finite games with $n$ players and $k_i$ strategies for player $i$, $i=1,\cdots,n$. Denote  $\kappa:=\prod_{i=1}^nk_i$.
\end{dfn}

Identify the strategy $j\in S_i$ with $\d_{k_i}^j$ called the vector form of strategy $j$. Then $c_i$ can be expressed algebraically as
\begin{align}\label{2.2.201}
c_i(x_1,\cdots,x_n)=V_i\ltimes_{j=1}^nx_j,
\end{align}
where $V_i$ is called the structure vector of payoff function $c_i$, $i=1,\cdots,n$.

\begin{dfn}\label{d2.2.2} \cite{mon96b} A finite game $G\in {\cal G}_{[n,k_1,\cdots,k_n]}$ is said to be a WPG, if there exists a function $P:S\ra \R$, called the potential function, and $n$ constants $w_i>0$, $i=1,\cdots,n$ called weights, such that
\begin{align}\label{2.2.3}
\begin{array}{l}
c_i(x_i,s_{-i})-c_i(y_i,s_{-i})=w_i\left(P(x_i,s_{-i})-P(y_i,s_{-i})\right),\\
\forall x_i,y_i\in S_i,\;\forall s_{-i}\in S_{-i}:=\prod_{j\neq i}S_j, ~i=1,\cdots,n,
\end{array}
\end{align}
If $w_i=1$, $i=1,\cdots,n$, $G$ is called a (pure) PG.
\end{dfn}

A repeated game is called an evolutionary game. If the strategies of player $i$ at time $t+1$, denoted by $x_i(t+1)$, depend only on the strategies $x_j(t)$ and payoffs $c_j(t)$ of players $j=1,\cdots,n$, it is called a Markoven type evolutionary game.

\begin{prp}\label{p2.2.3} \cite{che15} The strategy profile dynamics (SPD) of an Markoven evolutionary game can be expressed as
\begin{align}\label{2.2.4}
\begin{cases}
x_1(t+1)=f_1(x_1(t),\cdots,x_n(t))\\
\vdots\\
x_n(t+1)=f_n(x_1(t),\cdots,x_n(t)),\\
\end{cases}
\end{align}
where $f_i:\prod_{j=1}^n\D_{k_j}\ra \D_{k_i}$, $i=1,\cdots,n$ are determined by the strategy updating rule.
\end{prp}

Again using the vector expression of strategies (identify $x_i=j\in S_i$ with $\d_{k_i}^j$), the evolutionary dynamic equation (\ref{2.2.4}) can be expressed in its algebraic form as
\begin{align}\label{2.2.5}
x(t+1)=Lx(t),
\end{align}
where $x(t)=\ltimes_{i=1}^nx_i(t)$,  $L\in {\cal L}_{\kappa\times \kappa}$.

The SPD of a dynamic game is determined by the strategy updating rule. To simplify the statement, in this paper we assume the strategy updating rule is myopic best response arrangement (MBRA), that is,
\begin{align}\label{2.2.6}
x_i(t+1)=\underset{s_i\in S_i}{\mathrm{\argmax}}~c_i(s_i,s_{-i}(t)).
\end{align}

\section{Weighted Potential Games}

In this section we first review some results about finite PGs.  Extensions are done to make them applicable to WPGs. Finally, an algorithm is proposed for verifying WPGs and calculating their weights.

\subsection{Equation for a Weighted Potential Game}

The following proposition is based on  results in \cite{che14}. The extension has also been presented in {\cite{che16}.

\begin{prp}\label{p3.1.1} Assume $G\in {\cal G}_{[n;k_1,\cdots,k_n]}$.
\begin{itemize}
\item[(i)]~~$G$ is a WPG with weights $w_i$, $i=1,\cdots,n$, if and only if, the following potential equation (\ref{3.1.1}) has solution:
\begin{align}\label{3.1.1}
E_w\xi=B,
\end{align}
where
\begin{align}\label{3.1.2}
E_w=\begin{bmatrix}
-w_2E_1&w_1E_2&0&\cdots&0\\
-w_3E_1&0&w_1E_3&\cdots&0\\
\vdots&\vdots&~&~&\vdots\\
-w_nE_1&0&0&\cdots&w_1E_n
\end{bmatrix},
\end{align}
\begin{align}\label{3.1.3}
B=\begin{bmatrix}
\left(w_1V_2-w_2V_1\right)^T\\
\left(w_1V_3-w_3V_1\right)^T\\
\vdots\\
\left(w_1V_n-w_nV_1\right)^T\\
\end{bmatrix};
\end{align}
and
\begin{align}\label{3.1.4}
E_i=I_{\prod_{j=1}^{i-1}k_j}\otimes \J_{k_i}\otimes I_{\prod_{j=i+1}^{n}k_j},\quad i=1,\cdots,n.
\end{align}
\item[(ii)]~~ Let $\xi=(\xi_1^T,\xi_2^T,\cdots,\xi_n^T)^T$ be a solution of (\ref{3.1.1}), where $\xi^1_i\in \R^{\kappa/k_i}$. Then the game has its potential function as $P=V^P\ltimes_{j=1}^nx_j$, where
\begin{align}\label{3.1.5}
V^P=\frac{1}{w_i} \left[V_i-\xi_i^TE_i^T\right],\quad 1\leqs i\leqs n.
\end{align}
\end{itemize}
\end{prp}

\begin{rem}\label{r3.1.2}

\begin{itemize}
\item[(i)]
In equation (\ref{3.1.4}) we assume $I_{\emptyset}:=1$. That is,
$$
I_{\prod_{j=1}^0k_j}:=1; \quad I_{\prod_{j=n+1}^nk_j}:=1,
$$
which means that such factors do not exist.

\item[(ii)] (\ref{3.1.5}) can be used to calculate $V^P$ by choosing an arbitrary $i$ from $\{1,\cdots,n\}$.
\end{itemize}
\end{rem}

\subsection{Basis of Weighted Potential Games}

According to Proposition \ref{p3.1.1},  it is obvious that $G$ is a WPG with $\{w_i>0\;|\;1\leqs i\leqs n\}$, if and only if,
\begin{align}\label{3.2.1}
\begin{pmatrix}
w_1V_2^T-w_2V_1^T\\
w_1V_3^T-w_3V_1^T\\
\vdots\\
w_1V_n^T-w_nV_1^T
\end{pmatrix} =E_w\xi \in \Span(E_w).
\end{align}
Define $E^e=\begin{bmatrix}
w_1I_{\kappa}&0\\
0&E_w
\end{bmatrix}$, then (\ref{3.2.1}) can be rewritten as
\begin{align}\label{3.2.2}
\begin{pmatrix}
w_1V_1^T\\
w_1V_2^T-w_2V_1^T\\
w_1V_3^T-w_3V_1^T\\
\vdots\\
w_1V_n^T-w_nV_1^T
\end{pmatrix} =E^e\begin{bmatrix}V_1^T\\ \xi\end{bmatrix}\in \Span(E^e).
\end{align}
The left hand side of (\ref{3.2.2}) can further be expressed as
$$
C\begin{pmatrix}
V_1^T\\
V_2^T\\
\vdots\\
V_n^T
\end{pmatrix},
$$
where
$$
C=\begin{bmatrix}
w_1I_{\kappa}&0&\cdots&0\\
-w_2I_{\kappa}&w_1I_{\kappa}&\cdots&0\\
~&~&\ddots&~\\
-w_nI_{\kappa}&0&\cdots&w_1I_{\kappa}\\
\end{bmatrix}.
$$
It is easy to calculate that
$$
C^{-1}=\begin{bmatrix}
\frac{1}{w_1}I_{\kappa}&0&\cdots&0\\
\frac{w_2}{w_1^2}I_{\kappa}&\frac{1}{w_1}I_{\kappa}&\cdots&0\\
~&~&\ddots&~\\
\frac{w_n}{w_1^2}I_{\kappa}&0&\cdots&\frac{1}{w_1}I_{\kappa}\\
\end{bmatrix}.
$$
Next, we define
\begin{align}\label{3.2.3}
\begin{array}{ccl}
E^P_w&:=&w_1C^{-1}E^e\\
~&=&\begin{bmatrix}
w_1I_{\kappa}&0\\
\begin{array}{c}
w_2I_{\kappa}\\
w_3I_{\kappa}\\
\vdots\\
w_nI_{\kappa}
\end{array}&E_w
\end{bmatrix}\\
~&=&
\begin{bmatrix}
w_1I_{\kappa}&0&0&0&\cdots&0\\
w_2I_{\kappa}&-w_2E_1&w_1E_2&0&\cdots&0\\
w_3I_{\kappa}&-w_3E_1&0&w_1E_3&\cdots&~\\
~&~&~&~&\ddots&0\\
w_nI_{\kappa}&-w_nE_1&0&0&\cdots&w_1E_n\\
\end{bmatrix}\\
~&\in& {\cal M}_{n\kappa \times s},
\end{array}
\end{align}
where
\begin{align}\label{3.2.301}
s=\kappa+\frac{\kappa}{k_1}+\frac{\kappa}{k_2}+\cdots+\frac{\kappa}{k_n}.
\end{align}
Then we have
\begin{align}\label{3.2.4}
V^T_G:=\begin{pmatrix}
V_1^T\\
V_2^T\\
\vdots\\
V_n^T
\end{pmatrix}=C^{-1}E^e\begin{bmatrix}
V^T_1\\
\xi
\end{bmatrix}=\frac{1}{w_1}E^P_w\begin{bmatrix}
V^T_1\\
\xi
\end{bmatrix}.
\end{align}

\begin{rem} \label{r3.2.01} It is worth noting that even if $\xi$ has already been known we are still not able to determine $V_G$ completely by (\ref{3.2.4}). In fact, the condition for $G$ to be a WPG depends on only the differences $V_i-V_1$, $i=2,\cdots,n$. Hence $V_1$ is completely free. To determine $V_G$ by $\xi$ we can simply set $V_1=0$ (or any constant vector). This is important in numerical calculations.
\end{rem}

Denote the set of weighted potential games $G\in {\cal G}_{[n;k_1,\cdots,k_n]}$ by
$G\in {\cal G}^w_{[n;k_1,\cdots,k_n]}$. Then  the above argument yields the following result:

\begin{prp}\label{p3.2.1}  Given $G\in {\cal G}_{[n;k_1,\cdots,k_n]}$, then $G\in {\cal G}^w_{[n;k_1,\cdots,k_n]}$
  with $\{w_i|1\leqs i\leqs n\}$, if and only if,
\begin{align}\label{3.2.5}
V^T_G\in \Span(E^P_w).
\end{align}
Precisely speaking, there exists $\xi\in \R^{s-\kappa}$ such that (\ref{3.2.4}) holds.
\end{prp}

In the light of the argument in \cite{che14}, one sees easily that $E^P_w$ has codimension $1$ with respect to columns. Particularly, deleting the last column of $E^P_w$, denote what remains by
\begin{equation}\label{3.2.10}
	\tilde{E}^P_w=E^P_w\backslash\{\Col_s(E^P_w)\},
\end{equation}
then $\tilde{E}^P_w$ is of full column rank.

We also have the following

\begin{prp}\label{p3.2.2}
${\cal G}^w_{[n;k_1,\cdots,k_n]}$
is a vector subspace of ${\cal G}_{[n;k_1,\cdots,k_n]}\simeq \R^{n\kappa}$, satisfying
\begin{align}\label{3.2.6}
\dim\left( {\cal G}^w_{[n;k_1,\cdots,k_n]}\right)=\kappa+\dsum_{i=1}^n\frac{\kappa}{k_i}-1.
\end{align}
Moreover, $\Col (\tilde{E}^P_w)$ is its basis.
\end{prp}

Using least square approach, the following conclusion is obvious.

\begin{cor}\label{c3.2.3} Assume $G\in {\cal G}_{[n;k_1,\cdots,k_n]}$ with its structure vector $V_G$, and a set of weights $\{w_i\;|\;i=1,\cdots,n\}$ are given. The closest weighted potential game of $G$ having $\{w_i\}$ as its weights, denoted by $G_w$, has its structure vector as
\begin{align}\label{3.2.7}
V_{G_w}^T=\tilde{E}^P_wx,
\end{align}
where $\tilde{E}^P_w$ is defined in (\ref{3.2.10}), and
\begin{align}\label{3.2.8}
x=\left[(\tilde{E}^P_w)^T\tilde{E}^P_w\right]^{-1}(\tilde{E}^P_w)^TV_G^T
\end{align}

Moreover, since ${\cal G}_{[n;k_1,\cdots,k_n]}\simeq \R^{n \kappa}$, the Euclidean distance between $G$ and $G_w$ is
\begin{align}\label{3.2.9}
d(G,G_w)=\big\|V_G^T-V_{G_w}^T\big\|_2=\sqrt{(V_G-V_{G_w})(V_G^T-V_{G_w}^T)}.
\end{align}
\end{cor}

\subsection{Calculating Weights}

Without loss of generality, assume $w_1=1$. Then (\ref{3.1.2}) can be rewritten as
\begin{align}\label{3.3.1}
E_w=\left(-\begin{bmatrix}
E_1&0&\cdots&0\\
0&E_1&\cdots&0\\
~&~&\ddots&~\\
0&0&\cdots&E_1
\end{bmatrix}
\begin{bmatrix}
w_2\\w_3\\\vdots\\w_n
\end{bmatrix}
,\begin{bmatrix}
E_2&0&\cdots&0\\
0&E_3&\cdots&0\\
~&~&\ddots&~\\
0&0&\cdots&E_n
\end{bmatrix}\right).
\end{align}
Similarly, (\ref{3.1.3}) can be written as
\begin{align}\label{3.3.2}
B=-\begin{bmatrix}
(V_1)^T&0&\cdots&0\\
0&(V_1)^T&\cdots&0\\
~&~&\ddots&~\\
0&0&\cdots&(V_1)^T
\end{bmatrix}
\begin{bmatrix}
w_2\\w_3\\\vdots\\w_n
\end{bmatrix}
+\begin{bmatrix}
V_2^T\\
V_3^T\\
\vdots\\
V_n^T
\end{bmatrix}
\end{align}
Plugging (\ref{3.3.1})-(\ref{3.3.2}) into (\ref{3.1.1}) yields
\begin{align}\label{3.3.3}
\begin{array}{l}
-\left(I_{n-1}\otimes E_1\right)\begin{bmatrix}w_2\\ w_3\\ \vdots\\ w_n\end{bmatrix} \xi_1+
\begin{bmatrix}
E_2&0&\cdots&0\\
0&E_3&\cdots&0\\
~&~&\ddots&~\\
0&0&\cdots&E_n
\end{bmatrix}\begin{bmatrix}\xi_2\\ \xi_3\\ \vdots\\ \xi_n\end{bmatrix}\\
=-\left(I_{n-1}\otimes E_1\right)W_{[\kappa/k_1,n-1]}\xi_1\begin{bmatrix}w_2\\ w_3\\ \vdots\\ w_n\end{bmatrix}\\~~~+
\begin{bmatrix}
E_2&0&\cdots&0\\
0&E_3&\cdots&0\\
~&~&\ddots&~\\
0&0&\cdots&E_n
\end{bmatrix}\begin{bmatrix}\xi_2\\ \xi_3\\ \vdots\\ \xi_n\end{bmatrix}\\
=-\left(I_{n-1}\otimes V_1^T\right)\begin{bmatrix}w_2\\ w_3\\ \vdots\\ w_n\end{bmatrix}+
\begin{bmatrix}
V_2^T\\
V_3^T\\
\vdots\\
V_n^T
\end{bmatrix}.
\end{array}
\end{align}
Hence, we have
\begin{align}\label{3.3.4}
\begin{array}{l}
\left[\left(I_{n-1}\otimes E_1\right)W_{[\kappa/k_1,n-1]}\xi_1-\left(I_{n-1}\otimes V_1^T\right)\right]\begin{bmatrix}w_2\\ w_3\\ \vdots\\ w_n\end{bmatrix}
\\
=\begin{bmatrix}
E_2&0&\cdots&0\\
0&E_3&\cdots&0\\
~&~&\ddots&~\\
0&0&\cdots&E_n
\end{bmatrix}\begin{bmatrix}\xi_2\\ \xi_3\\ \vdots\\ \xi_n\end{bmatrix}-
\begin{bmatrix}
V_2^T\\
V_3^T\\
\vdots\\
V_n^T
\end{bmatrix}.
\end{array}
\end{align}
Next, we provide an algorithm to verify whether a finite game is a weighted potential one.

\begin{alg}\label{a3.3.1}
\begin{itemize}
\item Step 1
\begin{itemize}
\item 1-a:~Assume $w_i=w^0_i=1$,  $i=1,\cdots,n$. Using (\ref{3.2.8}), a least square solution of (\ref{3.1.1}) can be calculated as
    $$
    \xi^1=\begin{bmatrix}x\\0\end{bmatrix} =\begin{bmatrix}\xi^1_1\\ \xi^1_2\\ \vdots\\ \xi^1_n\end{bmatrix}\in \R^{\kappa_0},
    $$
    where
    $\kappa_0=\dsum_{i=1}^n\kappa/k_i$; $\xi^1_i\in \R^{\kappa/k_i}$, $i=1,\cdots,n$.

\item 1-b: If $\xi^1$ satisfies (\ref{3.1.1}), then the game is potential, we are done.

\item 1-c: Assume $\xi=\xi^1$. Use (\ref{3.3.4}) to find the least square solution $w^1$. Then go to next step.

\end{itemize}

\item Step k

\begin{itemize}

\item k-a: Setting $w=w^{k-1}$, which is obtained from Step $k-1$, calculate the least square solution $\xi^{k}$ by (\ref{3.2.8}).

\item k-b: Assume $\xi=\xi^{k}$. Use (\ref{3.3.4}) to find the least square solution $w^k$.

\item k-c: If $\|\xi^k-\xi^{k-1}\|_2<\epsilon$ and $\|w^k-w^{k-1}\|_2<\epsilon$, stop.
Use (\ref{3.2.4}) to calculate $V_{G_w}$ (keeping Remark \ref{r3.2.01} in mind), and then check whether $d(G,G_w)<\epsilon$, if ``yes", $G$ is WPG (within allowed error).

Otherwise, $G$ is not a WPG. (Refer to the following arguments about possible conclusions for this case.)

\end{itemize}
\end{itemize}
\end{alg}

\begin{rem} \label{r3.3.2}
\begin{itemize}
\item[(i)] Set $w=w^k$ and $\xi=\xi^{k-1}$, if (\ref{3.3.4}) is satisfied (within allowed error), then the given game is weighted potential.

\item[(ii)] Otherwise, Algorithm \ref{a3.3.1} provides a WPG closest to the given one. Then one can use (\ref{3.1.5}) to calculate the potential (Note that now $w_1=1$), which will be the potential of the  closest WPG.
\end{itemize}
\end{rem}

\begin{prp}\label{p3.3.3} If a given game $G$ is weighted potential, then algorithm \ref{a3.3.1} will converge to its weights, and (\ref{3.1.5}) will provide its potential.
\end{prp}

\noindent{Proof}. From the algorithm one sees easily that each step the square error is monotonically decreasing. Hence, eventually the error will achieve its minimum. When the original game is indeed a WPG, the square error will eventually arrive at zero, that is, we will have a true solution of (\ref{3.1.1}).
\hfill $\Box$

\begin{exa}\label{e3.3.4} Consider a game $G\in {\cal G}_{[2;2,3]}$, which has payoff bimatrix as shown in Table \ref{tb.3.3.1}.

\vskip 2mm

\begin{table}[!htb]
\centering\label{tb.3.3.1}
\caption{Payoff Bimatrix of Example \ref{e3.3.4}}
\vskip .25\baselineskip
\begin{tabular}{|c||c|c|c|}
\hline
$P_1\backslash P_2$&$1$&$2$&$3$\\
\hline
\hline
$1$&$(5,~0)$&$(1,~4)$&$(0,~2)$\\
\hline
$2$&$(2,~-2)$&$(-1,~2)$&$(1,~8)$\\
\hline
\end{tabular}
\end{table}

It is clear that
$$
\begin{array}{l}
V_1=[5,2,0,2,-1,1],\\
V_2=[0,4,2,-2,2,8].
\end{array}
$$
$$
\begin{array}{l}
E_1=\J_2\otimes I_3,\\
E_2=I_2\otimes \J_3.
\end{array}
$$

Applying Algorithm \ref{a3.3.1}, we have the following:

\begin{itemize}
\item Step 1-a: Assume $w^0_1=w^0_2=1$, the least square solution is:
$$
\xi^0_1=[3.3333,-3.6667,-5.6667]^T;~ \xi^0_2=[-2.3333,0]^T.
$$
\item Step 1-b: Check if $d(G,G_w)=0$, the answer in ``no". Then we continue.

\item Step 1-c: Use $\xi^0$ and formula (\ref{3.3.4}), an updated set of weights is obtained as
$w^1_1=1$ and $w^1_2=1.0443$.

\item Then we continue the iterations.
\end{itemize}

If we set $\epsilon=10^{-6}$, then after $214$ steps it turns out that
$$
w =2.0000,
$$
$$
\xi_1=[3.0000,-2.0000,-3.0000]^T; ~\xi_2=[-4.0000,0].
$$
And the structure vector of the potential function is
$$
V^P =[2.0000,4.0000,3.0000,-1.0000,1.0000,4.0000].
$$
We conclude that this game is a WPG with $w_1=1$, $w_2=2$.

\end{exa}

\section{Near Weighted Potential Games}

Since ${\cal G}^w_{[n;k_1,\cdots,k_n]}$ is not dense in ${\cal G}_{[n;k_1,\cdots,k_n]}$, one cannot always find a WPG close to a given finite game within an arbitrary distance; but still, we can consider the resemblance of their dynamics when two games are both repeated, which motivates the investigation of near weighted potential games (NWPGs).

\subsection{Verification of NWPGs}

Near potential games were firstly proposed by \cite{can13}. Its basic idea is: if a finite game is close to a PG, then the original game may have the same dynamic behavior as the potential one. Intuitively, a near potential game (NPG) comes from a PG with certain mild perturbations on its payoff functions. Based on the orthogonal decomposition of finite games, we propose an alternative way to construct near potential games: starting from an arbitrary finite game $G$, find its closest PG, denoted by $G_P$. Then verify whether $G$ and $G_P$ have the same strategy profile dynamics.  If ``yes", $G$ is called an NPG.

Note that when a set of weights is fixed, corresponding to this set we have also a similar orthogonal decomposition as follows \cite{wan17}:
\begin{align}\label{4.1.1}
{\cal G}=\rlap{$\underbrace{\phantom{\quad{\cal G}_w^{P_0}\quad\oplus\quad{\cal G}^N}}_{{\cal G}_w^P}$}\quad{{\cal G}^{P_0}_w}\quad\oplus\quad
\overbrace{{\cal G}^N\quad\oplus\quad{\cal G}_w^{H_0}}^{{\cal G}_w^H},
\end{align}
From now on we consider the WPG-based NWPGs. First we give a rigorous definition of it.

\begin{dfn} \label{d4.1.1} Given two finite evolutionary game $G$, $G'$,
\begin{itemize}
\item[(i)] $G$ and $G'$ are said to be evolutionary equivalent, if their strategy profile dynamics are identical.

\item[(ii)] If there exists an evolutionary (weighted) potential game $G^P$ ($G^P_w$) such that
$G$ and $G^P$ ($G^P_w$) are evolutionary equivalent, then $G$ is called an NPG (NWPG).
\end{itemize}
\end{dfn}

In the following we give an algorithm to verify if a finite game is an NWPG.

\begin{alg}\label{a.4.1.2}
\begin{itemize}
\item Step 1: Use Algorithm \ref{a3.3.1} to find the closest WPG, which has weights ~$w^*$ and least square solution  ~$\xi^*$ for the WPE.

\item Step 2: Set $\tilde{V}_1=V_1$, plug it into WPE (\ref{3.1.1}), solving out
\begin{align}\label{4.1.2}
\begin{bmatrix}
\tilde{V}_2^T\\
\tilde{V}_3^T\\
\vdots\\
\tilde{V}_n^T\\
\end{bmatrix}
=E_w\xi^*+\begin{bmatrix}
\tilde{V}_1^T\\
\tilde{V}_1^T\\
\vdots\\
\tilde{V}_1^T\\
\end{bmatrix}.
\end{align}
Then the game $\tilde{G}$ with $\tilde{V}_i$, $i=1,\cdots,n$ as its payoff functions and $w^*$ as its weights, is a WPG, which is closest to ~$G$.

\item Step 3: Compare the evolutionary dynamics of ~$G$ and ~$\tilde{G}$ to see whether they are evolutionary equivalent. If ``yes", $G$ is an NWPG.

\end{itemize}
\end{alg}

We consider the following example.

\begin{exa}\label{e.4.1.3} Given a finite game $G\in {\cal G}_{[3;2,2,3]}$. Assume the structure vectors of players' payoff functions are
$$
\begin{array}{l}
V_1=[2,3,-1,1,0,3,1,2,-2,2,2,3],\\
V_2=[-0.51,0.49,1,0,1,0,-1,-1.5,1.5,0.5,0.5,1];\\
V_3=[-2,0,2,8,10,6.1,2,4,6.1,2,6.1,-2].\\
\end{array}
$$
\begin{itemize}

\item[(i)] Find its closest PG:

Apply Algorithm \ref{a3.3.1}, and let $w_1=w_2=w_3$. The least square solution of PE is
$$
\xi_1=\begin{bmatrix}
	2.1163\\
	0.9163\\
	-3.6225\\
	0.3704\\
	-2.9796\\
	3.5092
\end{bmatrix},
~\xi_2=\begin{bmatrix}
	-0.5117\\
	-1.7867\\
	-0.5567\\
	-0.5067\\
	-3.5317\\
	0.6933
\end{bmatrix},
~\xi_3=\begin{bmatrix}
	-1.5300\\
	7.0000\\
	3.5033\\
	0
\end{bmatrix}.
$$
The structure vectors of payoff functions of the closest PG is
\begin{align}\label{4.1.3}
\begin{array}{ccl}
\tilde{V}_1&=&V_1,\\
\tilde{V}_2&=&[-0.6279,0.2971,2.0658,0.1179,1.1929,-1.0658,\\
~&~&-1.6229,-2.4479,2.3158,1.1229,1.4479,0.1842],\\
\tilde{V}_3&=&[-1.6463,0.5537,1.0925,7.6296,9.9796,6.4908,\\
~&~&2.3871,4.5871,5.1258,1.6296,4.9796,-0.5092].
\end{array}
\end{align}
It is easy to see that the error (distance between $G$ and $G^P$) is
\begin{align}\label{4.1.4}
d(V_G,V_{G^P})=3.3893,
\end{align}
which is large.

\item[(ii)] By Algorithm \ref{a3.3.1}, after 500 iterations, $G^P_w$ is obtained as follows:
\begin{itemize}
\item Its weights are
\begin{align}\label{4.1.5}
w_1=1,\quad w_2=0.5135,\quad w_3=2.0853.
\end{align}
\item The Structure vectors of its payoff functions are
\begin{small}
\begin{align}\label{4.1.6}
\begin{array}{ccl}
\tilde{V}_1&=&V_1,\\
\tilde{V}_2&=&[-0.4975,0.5027,1.0050,-0.0189,0.9837,0.0051,\\
~&~&-0.9999,-1.5082,1.4779,0.5064,0.5118,1.0119],\\
\tilde{V}_3&=&[-2.0080,-0.0074,2.0256,7.9984,10.0020,6.0893,\\
~&~&2.0049,4.0063,6.0786,2.0047,6.0991,-1.9936].
\end{array}
\end{align}
\end{small}
\item The error is
\begin{align}\label{4.1.7}
d(V_G,V_{G^P_w})=0.0579.
\end{align}
Compared with (\ref{4.1.4}) it is much smaller.

\item The structure vector of the potential function of $G^P_w$ is

\begin{align}\label{4.1.8}
\begin{array}{ccl}
\tilde{V}_P&=&[-0.9730,-0.0135,0.9709,-0.0402,\\
~&~&0.9220,-0.9566,-1.9730,-1.0135,\\
~&~&-0.0291,0.9598,2.9220,-0.9566],\\
\end{array}
\end{align}
\end{itemize}

\item[(iii)] Assume the MBRA is used as the strategy updating rule, it is easy to calculate that the strategy evolutionary dynamics of both $G$ and $G^P_w$ are the same (i.e. $G$ is an NWPG), which is
$$
\begin{array}{ccl}
x_1(t+1)&=&M_1x(t)\\
~&=&\d_2[1,1,1,2,2,1,1,1,1,2,2,1]x(t),\\
x_2(t+1)&=&M_2x(t)\\
&=&\d_2[2,2,1,2,2,1,2,2,1,2,2,1]x(t),\\
x_3(t+1)&=&M_3x(t)\\
~&=&\d_3[3,3,3,2,2,2,3,3,3,2,2,2]x(t),
\end{array}
$$
where $x_i(t)$ is the strategy of player $i$ at time $t$, $i=1,2,3$, and  $x(t)=\ltimes_{i=1}^3x_i(t)$. Finally, the overall strategy profile dynamics is
\begin{align}\label{4.1.9}
x(t+1)=Mx(t),
\end{align}
where
$$
\begin{array}{ccl}
M&=&M_1*M_2*M_3\\
~&=&\d_{12}[6,6,3,11,11,2,6,6,3,11,11,2].
\end{array}
$$

\end{itemize}
\end{exa}

\section{Application of NWPG to Game Theoretic Control}

The game theoretic control (GTC) is described in Figure \ref{Fig.5.1} according to \cite{gop11}.

\begin{figure}
\centering
\setlength{\unitlength}{0.8 cm}
\begin{picture}(9,7)\thicklines
\put(1,1){\line(1,0){7}}
\put(1,1){\line(1,1){2}}
\put(8,1){\line(-1,1){2}}
\put(3,3){\line(1,0){3}}
\put(3,3){\line(0,1){1}}
\put(3,4){\line(-1,1){2}}
\put(3,4){\line(1,0){3}}
\put(6,4){\line(0,-1){1}}
\put(6,4){\line(-1,0){3}}
\put(6,4){\line(1,1){2}}
\put(1,6){\line(1,0){7}}
\put(4.2,3.3){$\mbox{PG}$}
\put(3.4,5){$\mbox{Utility Design}$}
\put(3.2,2){$\mbox{Learning Design}$}
\end{picture}
\caption{Hourglass Architecture of Game Theoretical Control \label{Fig.5.1}}
\end{figure}
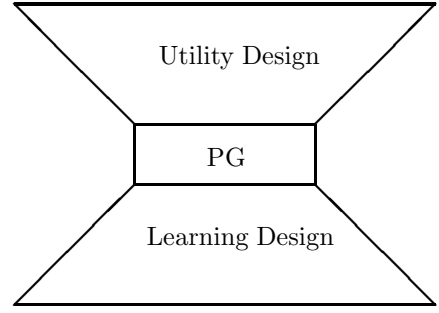

Roughly speaking, the optimization via game theoretic control can be described as follows: Consider a networked multi-agent system with $n$ agents. Assume there is a global objective function $J(x_1,\cdots,x_n)$, where $x_i$  is the action (or strategy) of the $i$-th agent. As is shown in Fig. \ref{Fig.5.1}, GTC approach consists of two major steps: (i) Design utility (or payoff) functions for each agents such that the overall system becomes a PG with $J$ as its potential function. (ii) Design a strategy updating rule (precisely, learning algorithm) such that when each agents are optimizing their own utility functions the system can converge to a Nash equilibrium, which is an optimal value of $J$ (a local one it might be). Note that, in general, since each agent can only obtain its neighbors' strategies, the learning algorithm must be based on local information.

As is discussed before, the application of PGs is very limited. In this section we provide a new method for game theoretic control by replacing PGs by NWPGs. Let $G\in {\cal G}_{[n;k_1,\cdots,k_n]}$ and the overall performance criterion $J(x_1,\cdots,x_n)$ be given. If there exists $G^P_w\in {\cal G}^w_{[n;k_1,\cdots,k_n]}$ with potential function $J$, which is evolutionary equivalent to $G$, then the GTC approach can be used for $G$. In other words, we consider structures of the top two parts of Figure \ref{Fig.5.1} and substitute them by Figure \ref{Fig.5.2}.


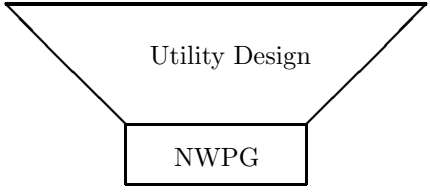
\begin{figure}
\centering
\setlength{\unitlength}{0.8 cm}
\begin{picture}(9,5)\thicklines
\put(3,1){\line(1,0){3}}
\put(3,1){\line(0,1){1}}
\put(3,2){\line(-1,1){2}}
\put(3,2){\line(1,0){3}}
\put(6,2){\line(0,-1){1}}
\put(6,2){\line(-1,0){3}}
\put(6,2){\line(1,1){2}}
\put(1,4){\line(1,0){7}}
\put(3.8,1.3){$\mbox{NWPG}$}
\put(3.4,3){$\mbox{Utility Design}$}
\end{picture}
\caption{NWPG-based Game Theoretic Control\label{Fig.5.2}}
\end{figure}

As for learning design, it is out of the scope of this paper. But we should emphasize that all the algorithms designed to realize the optimization for PG case are also applicable for NWPG case. We refer to \cite{fud98,mar12} for learning design.

To this end, we propose a method to reveal $G^P_w$, which is a WPG with potential function $J$ and is closest to $G$.

Rewrite (\ref{3.1.5}) in a matrix form as
\begin{align}\label{5.1}
\begin{array}{l}
\begin{bmatrix}
E_1&0&\cdots&0\\
0&E_2&\cdots&0\\
~&~&\ddots&~\\
0&0&\cdots&E_n
\end{bmatrix}
\begin{bmatrix}
\xi_1\\
\xi_2\\
\vdots\\
\xi_n
\end{bmatrix}
+\\
\begin{bmatrix}
(V^P)^T&0&\cdots&0\\
0&(V^P)^T&\cdots&0\\
~&~&\ddots&~\\
0&0&\cdots&(V^P)^T
\end{bmatrix}
\begin{bmatrix}
w_1\\
w_2\\
\vdots\\
w_n
\end{bmatrix}
=\begin{bmatrix}
V_1^T\\
V_2^T\\
\vdots\\
V_n^T
\end{bmatrix}.
\end{array}
\end{align}
For uniqueness of the WPG corresponding to $(\xi,w)$, set $\xi_n(\kappa)=0$ and $w_1=1$, define $\xi_0:=(\xi^T_n(1),\cdots,\xi^T_1(\kappa)$, $\cdots, \xi^T_n(1),\cdots,\xi^T_n(\kappa-1))^T$ and $w_0=(w_2,w_3,\cdots,w_n)$, where $\xi_i(j)$ means the $j$-th entry of $\xi_i$, and $E_n^0$ is obtained from $E_n$ by deleting its last column. Then (\ref{5.1}) can be expressed in a standard form as
\begin{align}\label{5.2}
E^0x=b,
\end{align}
where
$$
E^0=
\begin{bmatrix}
E_1&0&\cdots&0&0&0&\cdots&0\\
0&E_2&\cdots&0&0&(V^P)^T&\cdots&0\\
~&~&\ddots&~&\vdots&~&\ddots&~&~\\
0&0&\cdots&E_n^0&0&0&\cdots&(V^P)^T\\
\end{bmatrix},
$$
$$
x=(\xi^T_0,w^T_0)^T,
$$
and
$$
b=\begin{bmatrix}
(V_1-V^P)^T\\
V_2^T\\
\vdots\\
V_n^T
\end{bmatrix}.
$$
Then the least square solution of (\ref{5.2}) is
\begin{align}\label{5.3}
x=((E^0)^TE^0)^{-1}(E^0)^Tb.
\end{align}
Using (\ref{5.3}), we have
\begin{align}\label{5.4}
\begin{bmatrix}
\xi_1\\
\xi_2\\
\vdots\\
\xi_n\\
\end{bmatrix}
=\begin{bmatrix}
x(1)\\
\vdots\\
x(n\kappa-1)\\
0\\
\end{bmatrix};\quad
\begin{bmatrix}
w_1\\
w_2\\
\vdots\\
w_n\\
\end{bmatrix}
=\begin{bmatrix}
1\\
x(n\kappa)\\
\vdots\\
x(n\kappa-2)\\
\end{bmatrix}.
\end{align}
Using this set of solutions, $G^P_w$ can be constructed with the structure vectors $U_1,\cdots,U_n$ of its payoff functions as
\begin{align}\label{5.5}
\begin{array}{ccl}
	\begin{bmatrix}
		U_1^T\\
		U_2^T\\
		\vdots\\
		U_n^T
	\end{bmatrix}&=&\begin{bmatrix}
		E_1&0&\cdots&0\\
		0&E_2&\cdots&0\\
		~&~&\ddots&~\\
		0&0&\cdots&E_n
	\end{bmatrix}
	\begin{bmatrix}
		\xi_1\\
		\xi_2\\
		\vdots\\
		\xi_n
	\end{bmatrix}\\
	~&+&
	\begin{bmatrix}
		(V^P)^T&0&\cdots&0\\
		0&(V^P)^T&\cdots&0\\
		~&~&\ddots&~\\
		0&0&\cdots&(V^P)^T
	\end{bmatrix}
	\begin{bmatrix}
		w_1\\
		w_2\\
		\vdots\\
		w_n
	\end{bmatrix}.
\end{array}
\end{align}
If $G^P_w$ and $G$ are evolutionary equivalent, then $G^P_w$ can be used for game theoretic control.

We give an example to depict it.

\begin{exa}\label{e5.1} Consider a game $G\in {\cal G}_{[4;2,2,2,2]}$, where the payoff functions are expressed in their algebraic form as
$$
f_i=V_i\ltimes_{i=1}^4x_i,\quad i=1,2,3,4,
$$
with
$$
\begin{array}{ccl}
V_1&=&[310,217,108,158,131,260, 53, 29,\\
    ~&~&  88,172,283,235,314,  3,173,234],\\
V_2&=&[243,174, 80,120,103,203, 47, 25,\\
    ~&~&  72,143,235,192,251,  3,143,186],\\
V_3&=&[461,323,158,226,200,377, 77, 49,\\
    ~&~& 128,266,431,356,469,  3,257,351],\\
V_4&=&[367,262,131,190,159,298, 68, 40,\\
    ~&~&  97,206,334,282,378,  4,208,286].
\end{array}
$$
Assume the overall performance criterion is
$$
J(x_1,x_2,x_3,x_4)=V^P\ltimes_{i=1}^4x_i,
$$
where
$$
\begin{array}{ccl}
V^P&=&[300,210,100,150,130,250, 50, 30,\\
  ~&~&~ 80,170,280,230,310,  0,170,230].
\end{array}
$$
Applying Algorithm \ref{a3.3.1}, it is easy to verify that $G$ is neither a WPG nor a PG.

Next, we search for $G^P_w$. Using (\ref{5.3})-(\ref{5.5}), it is easy to calculate that
$$
\begin{array}{ccl}
U_1&=&[309.0,214.5,105.5,156.5,132.5,256.5, 53.0, 31.5,\\
    ~&~&~ 89.0,174.5,285.5,236.5,312.5,  6.5,173.0,231.5].
\end{array}
$$
$$
\begin{array}{ccl}
U_2&=&[241.0,172.5, 83.5,120.5,105.0,204.5, 43.5, 24.5,\\
    ~&~&~ 69.5,141.0,233.0,189.0,253.5,  5.0,145.0,189.0].
\end{array}
$$
$$
\begin{array}{ccl}
U_3&=&[460.6,319.8,158.4,229.2,198.9,379.2, 78.1, 46.8,\\
    ~&~&~128.4,265.7,430.6,356.3,468.8,  3.2,257.2,350.8].
\end{array}
$$
$$
\begin{array}{ccl}
U_4&=&[369.3,259.7,130.0,191.0,155.4,301.6, 66.2, 41.8,\\
    ~&~&~ 96.7,206.3,338.5,277.5,379.8,  2.2,207.1 280.2].
\end{array}
$$
It is easy to verify that $G^P_w$ is a WPG, with
$w=(1.0000,0.8004,1.5111,1.2184)$ and potential function as $J$.
Moreover, if MBRA is used for strategy updating rule, the strategy profile dynamics of $G$ and $G^P_w$ are the same, that is,
$$
x_i(t+1)=M_ix(t),\quad i=1,2,3,4,
$$
where
$$
\begin{array}{ccl}
M_1&=&\d_2[1,1,2,2,2,1,2,2,1,1,2,2,2,1,2,2],\\
M_2&=&\d_2[1,2,1,1,1,2,1,1,2,1,1,1,2,1,1,1],\\
M_3&=&\d_2[1,1,1,1,1,1,1,1,2,2,2,2,1,2,1,2],\\
M_4&=&\d_2[1,1,2,2,2,2,1,1,2,2,1,1,1,1,2,2].
\end{array}
$$
Hence $G$ and $G^P_w$ are evolutionary equivalent.
\end{exa}

\section{Conclusion}

The WPGs and NWPGs are investigated. First, an algorithm is provided to verify if a finite game is a WPG, and further determine the unknown weights. Meanwhile, when a game is not weighted potential, the algorithm can find its closest WPG. A finite game evolutionary equivalent to its closest WPG is called a NWPG, for which all main properties of a PG remain true. A criterion for NWPGs is given. In the end we proposed a method to find the closest WPG $G^P_w$ of a finite game $G$, which has a pre-assigned potential function; this result makes NWPGs applicable to game theoretic control when $G$ and $G^P_w$ are evolutionary equivalent.

\end{document}